\documentclass[aip,jcp,reprint,amsmath,amssymb,floatfix,superscriptaddress,groupedaddress]{revtex4-1}

\usepackage{graphicx}
\usepackage{dcolumn}
\usepackage{bm}
\usepackage{units}
\usepackage{epstopdf}

\def\cm{cm$^{-1}$}

\newcommand{\etal}{{\it et al.}}

\begin{document}


\title{Evolution of ferroelectricity in tetra\-thia\-fulvalene-p-chloranil as a function of pressure and temperature}
\author{Armin Dengl}
\author{Rebecca Beyer}
\author{Tobias Peterseim}
\author{Tomislav Ivek}
\author{Gabriele Untereiner}
\author{Martin Dressel}
\affiliation{1.\ Physikalisches Institut, Universit\"{a}t Stuttgart, Pfaffenwaldring 57, D-70550 Stuttgart, Germany}

\date{\today}

\begin{abstract}
The neutral-to-ionic phase transition in the mixed-stack charge-transfer complex
tetra\-thia\-fulvalene-p-chloranil (TTF-CA) has been studied by
pressure-dependent infrared spectroscopy up to $p=11$\,kbar and down to low
temperatures, $T=10$\,K. By tracking the C=O antisymmetric stretching mode of CA
molecules, we accurately determine the ionicity of TTF-CA in the
pressure-temperature phase diagram. At any point the TTF-CA
crystal bears only a single ionicity; there is no coexistence region or an exotic
high-pressure phase. Our findings shed new light on the role of electron-phonon
interaction in the neutral-ionic transition.
\end{abstract}


\maketitle

\section{Introduction}
While many of the exotic ground states investigated in condensed matter physics 
are only of academic interest, ferroelectricity has already realized
widespread applications for a century.\cite{Spaldin10, Nan08} Organic 
charge-transfer salts have an enormous potential in the field of ferroelectricity and
even multiferroicity;\cite{Brazovskii08,Horiuchi08,Lunkenheimer12,Dressel12} and
for many decades TTF-CA is considered the prime candidate.\cite{Torrance81} 
The mixed-stack compound is composed  
of the electron donor tetra\-thia\-fulvalene (TTF) and the
acceptor chloranil (CA), which makes the physical properties rather one
dimensional.\cite{LeCointe95} The charge transfer between TTF$^{(+\rho)}$ and
CA$^{(-\rho)}$ is strongly temperature-dependent and commonly described as
ionicity $\rho$. At ambient conditions the material is a paraelectric insulator
with a room-temperature $\rho = 0.2$;\cite{LemeeCailleau97} this phase is called
quasi-neutral. Upon cooling down, the ionicity gradually increases to $\rho =
0.3$ until the first-order transition takes place at $T_\mathrm{NI} = 81$\,K
characterized by a jump to $\rho \geq 0.5$ where TTF-CA enters the quasi-ionic
phase. The neutral-ionic transition (NIT) is accompanied by a Peiers-like
dimerization of the stacks leading to permanent dipoles and thus a ferroelectric
state.\cite{Okamoto91, LeCointe95, LemeeCailleau97, Kishida09, Masino200671, Girlando08} 
The ground state continues to be a subject of intense research because of many
outstanding questions and not yet completely understood properties such as
light-induced metastable states,\cite{Nagahori13} negative differential
resistance,\cite{Tokura88} anomalous dielectric response due to electronic
ferroelectricity,\cite{Okamoto91,Kobayashi12} a possibly multiferroic character
of the ionic phase,\cite{Giovannetti09} or an exotic high-pressure quasi-ionic
state.\cite{Masino07}

Ground states of many organic conductor are found to be highly sensitive to
hydrostatic pressure, and TTF-CA is no exception.\cite{Ishiguro98} The NIT in TTF-CA
can be driven either by pressure
($p$-transition)\cite{Torrance81PRL,Takaoka87,Moreac96} or temperature
($T$-transition).\cite{Torrance81} The pressure-driven transition at room
temperature has attracted enormous interest since its discovery almost three
decades ago.\cite{Tokura86, Mitani87} In contrast to the $T$-transition, the
$p$-transition does not exhibit a jump in ionicity but a gradual increase of $\rho$
which crosses 0.5 at about $p_{\rm NI} = 8.5$\,kbar.\cite{Matsuzaki05} This
leads to the open question of the role of the electron-phonon interaction in the
realization of the ionic state.

Optical spectroscopy is a powerful experimental method to determine the charge of
molecules and molecular materials.\cite{Dressel04,Drichko09,Girlando11,Sedlmeier12} 
Certain intra-molecular vibrations have a strong and in first approximation linear dependence
of their resonance frequencies on the molecular charge. If these
charge-sensitive vibrations are isolated in frequency, the molecular charge can
be easily and accurately estimated from infrared as well as Raman spectroscopy.
The findings coincide very well with theoretical calculations and alternative
methods such as NMR and x-ray scattering. In case the modes of interest
are not well-isolated in frequency or even overlap, the pressure and temperature
dependences may help to distinguish the charge-sensitive vibrations from other
features. In TTF-CA the most interesting vibration is the $\nu_{10}(b_{1u})$
stretching mode of the CA molecule which mainly involves the C=O
bond.\cite{Girlando83} Its frequency strongly depends on the charge and follows
the relation
\begin{equation}
\nu_{10}(\rho) = 1685\,{\rm cm}^{-1} - (160\,{\rm cm}^{-1} \times \rho) \quad .
\label{eq:nu10}
\end{equation}
Temperature-dependent experiments at ambient pressure clearly
show a jump in the resonance frequency at $T_\mathrm{NI}$. On the other hand,
pressure-dependent measurement at room temperature reveal only a gradual
evolution of the resonance frequency and suggest a different high-pressure
phase. There is an ongoing discussion whether already at $p_\mathrm{NI} =
8.5$\,kbar the system enters a state in which two sorts of CA-molecules with different
ionicities are present\cite{Masino04, Masino07} or whether a pressure of 16.5\,kbar and more
is necessary to induce the coexistence of quasi-ionic and quasi-neutral
phases.\cite{Matsuzaki05}

\begin{figure*}
\includegraphics[clip,width=1.0\linewidth]{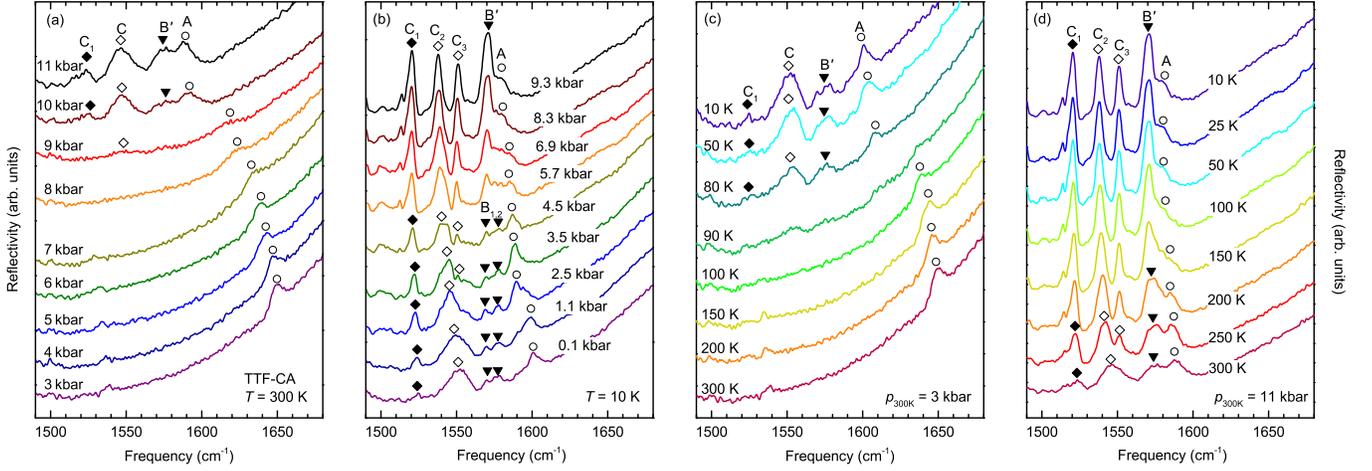}
\caption{\label{fig:1} Mid-infrared reflectivity of TTF-CA at different pressure
values and temperatures; the spectra are taken on powder samples and shifted for
clarity. Panels (a) and (b) show the pressure evolution at room temperature and
$T=10$\,K, respectively. The pressure values indicated in (b) refer to the actual
low-temperature pressure. Spectra belonging to the same room-temperature
pressure are plotted in the same color. Panels (c) and (d) exhibit the
temperature dependence starting with a pressure of 3 and 11\,kbar at room
temperature, respectively. The pressure decreases due to cooling from 3\,kbar (11\,kbar) at room temperature to 1.1\,kbar (10.1\,kbar) at 150\,K, 0.4\,kbar (9.8\,kbar) at 80\,K, down to 0.1\,kbar (9.6\,kbar) at 10\,K. The A band (marked with circles) is assigned
to the charge-sensitive C=O antisymmetric vibration of the CA molecule
$\nu_{10}(b_{1u})$. The B$^\prime$, B$_1$, B$_2$ and the C bands
(marked with triangles and rhombi) are discussed in the text. The appearance of
these bands indicates that the NIT happens at 8.5\,kbar for $T=300$\,K. At the
lowest measured temperature of 10\,K, the system is in the quasi-ionic phase for
all pressures. Upon cooling at 3\,kbar (room temperature pressure) the NIT is
marked by the appearance of B$^\prime$ and C bands below 90\,K. For a
pressure of 11\,kbar, TTF-CA is in the quasi-ionic phase even at $T=300$\,K.
}
\end{figure*}

Previous optical experiments of TTF-CA under pressure could give only an
incomplete picture since they were strongly limited in
temperature,\cite{Takaoka87} hence we investigated the neutral-ionic transition
in a wide temperature range down to $T=10$\,K and up to a pressure of 11\,kbar.
The ionicity of TTF-CA was extracted from the $\nu_{10}(b_{1u})$ stretching
vibration of CA for all measured pressures and temperatures. This gives some
insight concerning the coexistence of different ionicities at high pressures and
temperatures.

\section{Experimental}
Optical reflectivity measurements of TTF-CA samples were performed 
in the mid-infrared frequency range
from 500 to 8000\,\cm{} utilizing a Fourier-transform infrared spectrometer of
the type Bruker IFS 66v/S equipped with a CryoVac He-flow cryostat which could
reach temperature as low as 10\,K. In order to investigate the optical
properties under hydrostatic pressure, we employed a piston cylinder cell CC33
of the Institute of High-Pressure Physics, Russian Academy of Sciences, and
Daphne 7373 as pressure medium. The sample was directly mounted onto the type
IIa diamond window; hence, we probe the reflectivity off the diamond-sample
interface. In the frequency range between 1800 and 2700\,\cm{} the strong
multi-phonon absorption in diamond prevents reliable measurements. Pressure of
up to 11\,kbar was applied at room temperature and the cell closed. When the
copper beryllium cell is cooled down, the actual pressure is significantly reduced due
to thermal contraction. Hence we thoroughly determined the
temperature-dependent pressure loss beforehand, using ruby fluorescence
spectroscopy. With this pressure calibration, the real low-temperature
pressure is known within an error of $\pm 1$\,kbar.\cite{Beyer} The pressure
cell with the mounted sample was not opened until the measurements were
completed at all desired temperatures and pressure values.

Fresh TTF-CA single crystals were grown by plate sublimation method.\cite{Demianets80}
The obtained specimens, however, were too thin for optical measurement and
partially transparent in the mid-infrared frequency range. Therefore, the
measurements were performed on 1\,mg of powdered TTF-CA crystals that were
finely ground and firmly pressed. Obviously, we do not gain information on the
optical anisotropy and also the absolute conductivity values are not
particularly meaningful. With this in mind we analyzed the measured reflectivity spectra directly by a fit to a sum of Lorentz terms \cite{DresselGruner02} with additional Fano contributions in order to better describe coupled vibrational features.\cite{Fano61,Damascelli97,Sedlmeier12} Here the contribution of a
single vibrational model to the real and imaginary part of optical conductivity
reads
\begin{subequations}
\label{eq:Fano}
\begin{eqnarray}
\sigma_1^{\rm Fano}(\nu) &=&
\sigma_0\frac{\gamma\nu\left[\gamma\nu(q^2-1)+2q(\nu^2-\nu_0^2)\right]}
{(\nu^2-\nu_0^2)^2+\gamma^2\nu^2} \quad ,
\label{eq:Fano1}\\
\sigma_2^{\rm Fano}(\nu) &=&
\sigma_0\frac{\gamma\nu\left[(q^2-1)(\nu^2-\nu_0^2)-2\gamma\nu\right]}
{(\nu^2-\nu_0^2)^2+\gamma^2\nu^2} \quad ,
\label{eq:Fano2}
\end{eqnarray}
\end{subequations}
where $\nu_0$ is the resonance frequency, $q$ the phenomenological coupling
parameter, and $\sigma_0$ the amplitude of the Fano contribution. The linewidth
$\gamma$ relates to the damping constant $\tau$ as $\gamma=1/(2\pi c \tau)$.

\section{Results}

In Fig.\ \ref{fig:1}(a) the room temperature reflectivity spectra of powdered
TTF-CA crystals are plotted for the spectral range around the $\nu_{10}(b_{1u})$
mode. The increase of the reflectivity with frequency is an artifact caused by
the multi-phonon absorption of the diamond window that becomes dominant above
1700\,\cm; here we focus on the vibrational features seen as small but distinct
bands on top of it. At ambient condition and low pressure values, we identify a
single band around 1650\,\cm{} that is assigned to the $\nu_{10}(b_{1u})$ mode
of the C=O stretching vibration (labeled as A band). With increasing pressure
this band shifts to lower frequencies, and broadens slightly. Above 9\,kbar
additional features appear: the B$^\prime$ band at 1575\,\cm{} together with two C bands
at lower frequencies. As discussed in more details below, these bands are due to
symmetric vibrations that become infrared-active by the interaction with
electrons, the so-called electron-molecular vibration (emv)
coupling.\cite{Dressel04} At the pressure-induced neutral-ionic transition
$p_\mathrm{NI}=8.5$\,kbar the charge transfer increases and dimerization of the
TTF-CA stacks is modified; hence the emv-coupled modes become
pronounced.\cite{Girlando83} Our findings agree well with previous
pressure-dependent measurements at room temperature.

The temperature dependence of the reflectivity is displayed by
Figs.\ \ref{fig:1}(c) and (d) for low and high pressure, respectively. Note that
the indicated pressure values correspond to the pressure applied at $T=300$\,K;
it decreases with cooling. In Fig.\ \ref{fig:1}(c) it is seen that starting with
$p=3$\,kbar at room temperature, the $\nu_{10}(b_{1u})$ mode (A band) shifts
only weakly with cooling. The NIT is evident by a jump to lower frequencies at
$T_\mathrm{NI}\approx 80$\,K. Additional B$^\prime$ and C bands appear in the quasi-ionic
phase as observed for the $p$-driven transition [Fig.\ \ref{fig:1}(a)].

Fig.\ \ref{fig:1}(b) displays the pressure dependence of the bands of interest at
$T=10$\,K, the lowest temperature measured. The pressure values shown are
corrected for the pressure loss. Since the A band is always observed below
1605\,\cm, which corresponds to the threshold ionicity of 0.5, the system
remains in the quasi-ionic state for any pressure. Accordingly, the emv-coupled
B$^\prime$ and C bands can be identified at low temperatures for any pressure applied. The
B$^\prime$ further becomes resolved into two vibrations labeled B$_1$ and B$_2$ at lower
pressures. On the other hand, the C$_3$ band seems to merge with the C$_2$ at
low pressure. Similarly, the two modes cannot be distinguished in the room
temperature spectra even at high pressure, as seen in Fig.\ \ref{fig:1}(d).

In Fig.\ \ref{fig:2} the vibrational bands of TTF-CA found at lower frequencies
are displayed for different pressure values; the spectra are taken at $T=300$
and 10\,K, as indicated. The two asymmetric modes around 1110\,\cm{} and 1125\,\cm{} are
assigned to the $\nu_{15}(b_{1u})$ vibration of TTF and the $\nu_{11}(b_{1u})$
mode of CA, respectively. The assignment of the symmetric bands at 910\,\cm{}, 980\,\cm{} and
1275\,\cm{} is not that clear. The broad band at 980\,\cm{} is most likely due to
the $\nu_3(a_{g})$ vibration of CA,\cite{Girlando83} but the origin of the peaks
at high pressures and low temperatures [top curve of Fig.\ \ref{fig:2}(b)]
remains unresolved. All four vibrations are emv-coupled. Somewhat unexpectedly,
they all sharpen significantly under pressure at all temperatures.

\section{Discussion}
In Fig.\ \ref{fig:1} we have noticed that at high pressure values and at low
temperatures a new set of modes appear, labeled as B$^\prime$ and C bands.
Since they are not prominent in the neutral phase, it is safe to conclude that
these bands are infrared silent modes. The C$_2$ and C$_3$ bands
become infrared-active only by emv-coupling and hence are symmetric modes,
probably the $\nu_2(a_{g})$ mode of TTF and the $\nu_1(a_{g})$ mode of the CA
molecule, respectively.\cite{Girlando83} The C$_1$ mode is the only one which
can also be identified in the quasi-neutral phase, and thus we assign to the
asymmetric $\nu_{14}(b_{1u})$ mode of TTF.

\begin{figure}
\includegraphics[clip,width=1.0\linewidth]{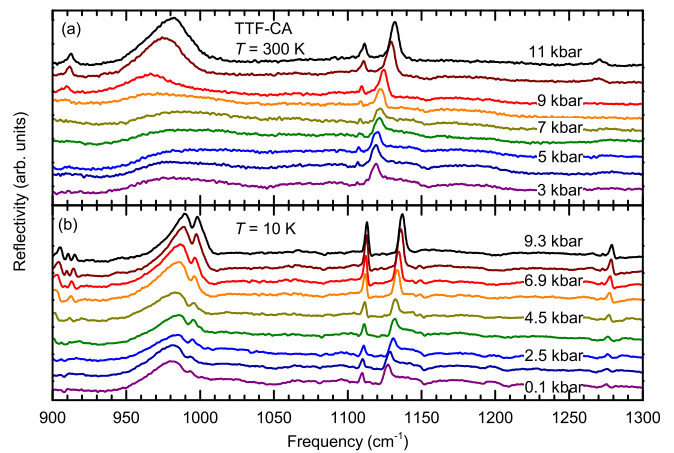}
\caption{\label{fig:2} Pressure-evolution of the low-frequency reflectivity
spectra of TTF-CA spectra measured at (a) room temperature and (b) $T=10$\,K.
For the assignment of the different bands see text. Note, that all features
become sharper with increasing pressure.}
\end{figure}

The assignment of the B$^\prime$ band around 1575\,\cm{} is not straightforward
as it contains two components, one of which was controversially discussed in
previous publications.\cite{Masino07,Matsuzaki05} In single-crystal measurements
a mode appears at 1570\,\cm{} below the transition temperature, for light
polarized parallel to the stacking direction. This mode can be unambiguously
assigned to the symmetric $\nu_2(a_g)$ vibration of CA. While the mode 
is present only in the ionized state, its resonance frequency does not depend on the
ionicity of CA.\cite{Girlando83} Driving TTF-CA through the $p$-dependent
transition at room temperature leads to the rise of an additional feature at
1580\,\cm{} above the critical pressure. This band was observed in
single-crystal measurements by several groups,\cite{Masino07, Matsuzaki05} but
its origin is still under discussion (it is commonly called B band). Masino
\etal{}\ link it to the $\nu_{10}(b_{1u})$ of CA; they also assign their A band
to the same CA vibration.\cite{Masino07} Accordingly, they claim the appearance
of a second ionicity and thus a new state in which two 
different sorts of molecules with distinct ionicities coexist. 
Their assignment is mostly based on the fact that this band is only
detected in the ionized state and that previous measurements and calculations do
not predict other molecular vibrations in this frequency range.\cite{Girlando83}
On the other hand, Matsuzaki \etal{}\ assign the same vibration to an
infrared-active mode, as they are able to identify the B band also in the
neutral state thanks to measurements of superior sensitivity.\cite{Matsuzaki05} No
matter whether one follows their assignment, it is important to notice that the
mode exists in the neutral state and that its frequency does not depend on the
ionicity.

At low temperatures and low pressures the two contributions to the B$^\prime$
band can be resolved and assigned: the lower-frequency B$_1$ band to the
$\nu_2(a_g)$ vibration and the B$_2$ band to the other vibration discussed
above. Applying a careful fitting procedure, we can further resolve the two
vibrations for higher pressures at the lowest temperature. Thus we conclude that
the high-temperature B$^\prime$ band consists of two components, one being
$\nu_2(a_g)$ of CA and the other an infrared-active mode of unclear assignment,
both of which are not charge sensitive. 
It is well-established that at ambient pressure the ionic
phase consists of only one sort of TTF-CA molecule with a single well-defined
ionicity. From Figs.\ \ref{fig:1}(b) and (d) it is clearly seen that the
B$^\prime$ band does exist at low temperatures for all pressure values measured,
and that it changes smoothly through cooling at high pressures; we see no
evidence of an extra $\nu_{10}(b_{1u})$ component. Within our temperature and
pressure range the high-pressure phase comprises only one type of CA molecule. Table \ref{tab:modes} summarizes the final mode assignment.

\begin{table} 
\centering
\caption{Assignment of select observed features to vibrational modes of TTF or CA.}
\begin{tabular*}{\linewidth}{@{\extracolsep{\fill}}clcccl}
\hline\hline
Label & Vibrational mode && Label & Vibrational mode \\
\hline
A & $\nu_{10}(b_{1u})$ of CA && C$_1$ & $\nu_{14}(b_{1u})$ of TTF \\
B$_1$ & $\nu_2(a_g)$ of CA && C$_2$ & $\nu_2(a_g)$ of TTF \\
B$_2$ & unclear assignment, see text && C$_3$ & $\nu_1(a_g)$ of CA \\

\hline\hline
\end{tabular*}
\label{tab:modes}
\end{table}

According to the above arguments there is no ground left to claim a non-dimerized
quasi-ionic phase of TTF-CA that exists at room temperature above 9\,kbar. 
In TTF-CA the emv-coupled modes are activated by the stack
dimerization.\cite{Girlando83} A non-dimerized quasi-ionic state should show
fewer, or different, emv-coupled modes compared to the regular state. However,
Figs.\ \ref{fig:1}(a) and \ref{fig:2}(a) provide strong evidence that the number
of emv-coupled vibrations does not change with increasing pressure; the same is
true for the high-pressure (11\,kbar) spectra with varying temperature, as shown
by Fig.\ \ref{fig:1}(d). These results rule out a non-dimerized quasi-ionic
phase of TTF-CA above 9\,kbar.

\begin{figure}
\includegraphics[clip,width=1.0\linewidth]{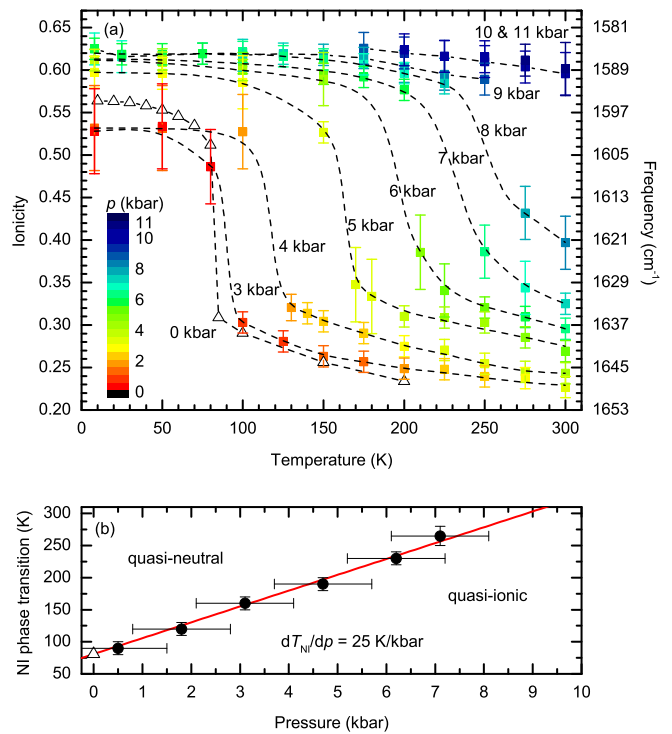}
\caption{\label{fig:3}(a) Temperature dependence of the ionicity $\rho$ of
TTF-CA at different hydrostatic pressure. The labels designate the value of
pressure at room temperature, while the lines follow separate cooling runs from
room temperature to 10\,K. The color of the points gives the corrected pressure
as shown by the legend. The empty triangles represent separate ambient-pressure
measurements on a single crystal. The quasi-ionic phase is defined by an
ionicity $\rho>0.5$. (b) Phase diagram of TTF-CA as obtained from our optical
measurements. The transition temperature under applied pressure (circle)
increases by 25\,K/kbar, starting from the 81\,K at ambient pressure
(triangle). Note that the pressure values shown here are corrected for the losses by thermal contraction.}
\end{figure}

From our experiments we can can safely conclude that TTF-CA crystals have only one
ionicity within our temperature and pressure ranges. From the advanced
fit of the reflectivity spectra with Fano terms for vibrational features we obtain the
resonance frequency of the $\nu_{10}(b_{1u})$ feature and calculate
the ionicity of TTF-CA at all temperatures and pressure values, according to
Eq.\ (\ref{eq:nu10}). At low pressures [cf.\ Fig.\ \ref{fig:1}(b)] and close
to the transition temperature [cf.\ Fig.\ \ref{fig:1}(c), 90\,K] the A band is
somewhat smeared out, which makes the estimation of the resonance frequency very
difficult in these regions. Accordingly, certain points in the phase diagram
contain rather large error bars. Fig.\ \ref{fig:3}(a) exhibits the ionicity of
TTF-CA as a function of temperature for different pressure runs. The corrected
pressure is displayed as the color of individual points. Even with the rough
temperature steps chosen in these optical experiments ($\Delta T \geq 10$\,K),
it can be seen that the transition gets broader with increasing pressure.
Surprisingly, in the quasi-neutral state both pressure and temperature have a
significant effect on the ionicity, while the ionicity hardly changes anymore
once the quasi-ionic phase is stabilized; neither with pressure nor with
temperature.

Based on our extensive optical investigations we can construct a revised phase
diagram of TTF-CA as a function of pressure and temperature, displayed in 
Fig.\ \ref{fig:3}(a). As our findings do
not give any indications for an exotic phase above 9\,kbar, only two different
phases are shown in Fig.\ \ref{fig:3}(b). The temperature of the NIT has a
linear dependence on pressure with a slope of 25\,K/kbar; note that we refer to
the actual pressure present at low temperatures. Extrapolation of our data
reveal that at room temperature the dimerized-stack, quasi-ionic phase sets in
at about $p_\mathrm{NI}=8.5$\,kbar. This optical phase diagram is in good
agreement with previous pressure-dependent
studies.\cite{Takaoka87,LemeeCailleau92,Mitani87} We do not observe a critical
pressure below which the NIT-temperature levels off at $T_\mathrm{NI}= 81$\,K.
This contradicts previous dc transport measurements of Mitani
\etal,\cite{Mitani87} who report a pressure-independent transition
temperature. This discrepancy at low-temperatures can be explained by the precaution
we have taken to carefully include the pressure loss with cooling in the entire
temperature range. The pressure loss in piston cylinder cells generally becomes
stronger at lower temperatures and is especially important at low pressures: for
instance, Fig.\ \ref{fig:1}(b) demonstrates how the pressure in the cell is reduced
from 3\,kbar at room temperature to only 0.1\,kbar at 10\,K.

There is another observation we would like to draw the attention to.
Usually vibrational modes become sharper upon cooling, 
as a consequence of thermal broadening. 
They also are smeared out when pressure is applied
because the phonon density of states spreads out with pressure, 
leading to a larger probability for phonon-phonon scattering and a widening of the vibration lines. 
In the present case of TTF-CA, however, our pressure- and temperature-dependent
measurements reveal a rather unexpected behavior. As shown in Figs.\ \ref{fig:1}(a),
(b) and \ref{fig:2}, the B$_1$, C$_1$ and C$_3$ bands as well as the lower-frequency
bands around 980, 1110, and 1250\,\cm{} clearly become more pronounced with pressure.
One might argue that the phonon-phonon interaction does not increase with
pressure, thus allowing the modes to narrow even though the phonon density of
states becomes wider. However, the mode narrowing is also seen at
temperatures as low as 10\,K where we do not expect the phonon-phonon scattering
to play any significant role.

\begin{figure}
\includegraphics[clip,width=1.0\linewidth]{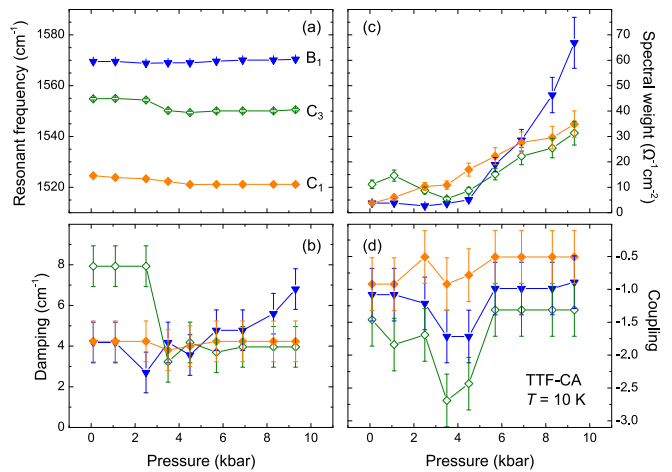}
\caption{\label{fig:4} Pressure dependence of the mode parameters obtained from
the Fano fit [Eq.\ (\ref{eq:Fano})] of select vibrational features of TTF-CA at
10\,K: (a) resonant frequency, (b) the damping, (c) spectral
weight (see text), and (d) the Fano coupling parameter. The error bars
indicate the uncertainty of fits to the reflectivity spectra of powder.}
\end{figure}

For a more quantitative analysis of these pressure-dependent modes we 
apply a Fano fit according to Eq.\ (\ref{eq:Fano}). 
It is instructive to represent the mode strength
of each Fano contribution by spectral weight $\int|\sigma_1(\nu)|\mathrm{d}\nu$
(the absolute value has to be taken because the Fano contribution to
conductivity can also be negative in sign). Fig.\ \ref{fig:4} displays the
low-temperature parameters of the B$_1$, C$_1$ and C$_3$ bands as a function of
pressure. For clarity's sake we leave out the low-frequency vibrational bands
which become stronger with pressure in a similar manner, as well as the A and
C$_2$ bands which do not show a clear trend with pressure. It is apparent
that for the modes shown in Fig.\ \ref{fig:4} the spectral weight increases
with pressure, especially above 4\,kbar.
This further corresponds with a dip in the coupling constant and
a change of damping in C$_3$ around 4\,kbar, the origin of which is as of yet not understood. The vibrations in question include some emv-coupled
as well as infrared-active modes.
As pressure is applied, the modes become more pronounced because the ionicity
is not independent on pressure. As demonstrated by
Fig.\ \ref{fig:3}(a), the low-temperature ionicity $\rho$ gradually increases with
pressure. Certain molecular vibrations depend on ionicity, and the coupling can change not only their resonant frequencies but also other parameters, including strength.
In the ionic phase of TTF-CA the latter is particularly noticable
for some CA modes [the B$_1$ band, $\nu_2(a_g)$; the C$_3$ band, $\nu_1(a_g)$; the
$\nu_{11}(b_{1u})$ around 1125\,\cm{}; the 980\,\cm{} band which is most likely
the $\nu_3(a_g)$ of CA] and TTF modes [the C$_1$ band, $\nu_{14}(b_{1u})$;
the $\nu_{15}(b_{1u})$ around 1110\,\cm{}].

\section{Conclusion}
We have performed comprehensive reflectivity measurements on TTF-CA under
hydrostatic pressure up to 11\,kbar in the temperature range from 300\,K down to
10\,K. By evaluating the charge-sensitive $\nu_{10}(b_{1u})$ vibrational mode of
the CA molecule, we determined the ionicity of the system as a function of both
temperature and pressure; hence we could compose a revised phase diagram of
TTF-CA which connects the temperature-driven neutral-ionic transition with a critical
temperature of $T_\mathrm{NI}=81$\,K and the pressure-driven transition with a
critical pressure of $p_\mathrm{NI}=8.5$\,kbar. By carefully comparing the
pressure-dependent spectra at high and low temperatures with the
temperature-dependent spectra at high and low pressures, we can exclude an
exotic high-pressure phase in the measured region. In contrast to previous
suggestions,\cite{Masino04,Masino07} the phase diagram constructed
from our optical measurements contains only two phases: the quasi-neutral and
one quasi-ionic phase. The neutral-to-ionic transition temperature increases
linearly with pressure by 25\,K/kbar. We additionally point out the atypical pressure dependence of the vibrational mode strength in the quasi-ionic phase, most likely due to the coupling to ionicity.

\begin{acknowledgments}
We wish to thank A. Girlando and M. Masino for enlightening discussions.
This work was supported by the Deutsche Forschungsgemeinschaft (DFG).
T.P. acknowledges the support of Carl Zeiss Stiftung.
\end{acknowledgments}


\begin{thebibliography}
\expandafter\ifx\csname natexlab\endcsname\relax\def\natexlab#1{#1}\fi
\expandafter\ifx\csname bibnamefont\endcsname\relax
  \def\bibnamefont#1{#1}\fi
\expandafter\ifx\csname bibfnamefont\endcsname\relax
  \def\bibfnamefont#1{#1}\fi
\expandafter\ifx\csname citenamefont\endcsname\relax
  \def\citenamefont#1{#1}\fi
\expandafter\ifx\csname url\endcsname\relax
  \def\url#1{\texttt{#1}}\fi
\expandafter\ifx\csname urlprefix\endcsname\relax\def\urlprefix{URL }\fi
\providecommand{\bibinfo}[2]{#2}
\providecommand{\eprint}[2][]{\url{#2}}

\bibitem[{\citenamefont{Nicola A.~Spaldin and Ramesh}(2010)}]{Spaldin10}
\bibinfo{author}{\bibfnamefont{S.-W.~C.} \bibnamefont{Nicola A.~Spaldin}}
  \bibnamefont{and} \bibinfo{author}{\bibfnamefont{R.}~\bibnamefont{Ramesh}},
  \bibinfo{journal}{Phys. Today} \textbf{\bibinfo{volume}{63}},
  \bibinfo{pages}{38} (\bibinfo{year}{2010}),
  \urlprefix\url{http://scitation.aip.org/content/aip/magazine/physicstoday/article/63/10/10.1063/1.3502547}.

\bibitem[{\citenamefont{Nan et~al.}(2008)\citenamefont{Nan, Bichurin, Dong,
  Viehland, and Srinivasan}}]{Nan08}
\bibinfo{author}{\bibfnamefont{C.-W.} \bibnamefont{Nan}},
  \bibinfo{author}{\bibfnamefont{M.~I.} \bibnamefont{Bichurin}},
  \bibinfo{author}{\bibfnamefont{S.}~\bibnamefont{Dong}},
  \bibinfo{author}{\bibfnamefont{D.}~\bibnamefont{Viehland}}, \bibnamefont{and}
  \bibinfo{author}{\bibfnamefont{G.}~\bibnamefont{Srinivasan}},
  \bibinfo{journal}{J. App. Phys.} \textbf{\bibinfo{volume}{103}},
  \bibinfo{pages}{031101} (\bibinfo{year}{2008}),
  \urlprefix\url{http://scitation.aip.org/content/aip/journal/jap/103/3/10.1063/1.2836410}.

\bibitem[{\citenamefont{Brazovskii}(2008)}]{Brazovskii08}
\bibinfo{author}{\bibfnamefont{S.~A.} \bibnamefont{Brazovskii}},
  \emph{\bibinfo{title}{The Physics of Organic Superconductors and Conductors}}
  (\bibinfo{publisher}{Springer Verlag}, \bibinfo{address}{Berlin},
  \bibinfo{year}{2008}).

\bibitem[{\citenamefont{Horiuchi and Tokura}(2008)}]{Horiuchi08}
\bibinfo{author}{\bibfnamefont{S.}~\bibnamefont{Horiuchi}} \bibnamefont{and}
  \bibinfo{author}{\bibfnamefont{Y.}~\bibnamefont{Tokura}},
  \bibinfo{journal}{Nat. Mater.} \textbf{\bibinfo{volume}{7}},
  \bibinfo{pages}{357} (\bibinfo{year}{2008}), ISSN \bibinfo{issn}{1476-1122},
  \urlprefix\url{http://dx.doi.org/10.1038/nmat2137}.

\bibitem[{\citenamefont{Lunkenheimer et~al.}(2012)\citenamefont{Lunkenheimer,
  M\"{u}ller, Krohns, Schrettle, Loidl, Hartmann, Rommel, de~Souza, Hotta,
  Schlueter et~al.}}]{Lunkenheimer12}
\bibinfo{author}{\bibfnamefont{P.}~\bibnamefont{Lunkenheimer}},
  \bibinfo{author}{\bibfnamefont{J.}~\bibnamefont{M\"{u}ller}},
  \bibinfo{author}{\bibfnamefont{S.}~\bibnamefont{Krohns}},
  \bibinfo{author}{\bibfnamefont{F.}~\bibnamefont{Schrettle}},
  \bibinfo{author}{\bibfnamefont{A.}~\bibnamefont{Loidl}},
  \bibinfo{author}{\bibfnamefont{B.}~\bibnamefont{Hartmann}},
  \bibinfo{author}{\bibfnamefont{R.}~\bibnamefont{Rommel}},
  \bibinfo{author}{\bibfnamefont{M.}~\bibnamefont{de~Souza}},
  \bibinfo{author}{\bibfnamefont{C.}~\bibnamefont{Hotta}},
  \bibinfo{author}{\bibfnamefont{J.~A.} \bibnamefont{Schlueter}}, \bibnamefont{and}
  \bibinfo{author}{\bibnamefont{M.} \bibnamefont{Lang}}, \bibinfo{journal}{Nat. Mater.}
  \textbf{\bibinfo{volume}{11}}, \bibinfo{pages}{755} (\bibinfo{year}{2012}),
  \urlprefix\url{http://www.nature.com/nmat/journal/v11/n9/full/nmat3400.html}.

\bibitem[{\citenamefont{Dressel et~al.}(2012)\citenamefont{Dressel, Dumm,
  Knoblauch, K\"{o}hler, Salameh, and Yasin}}]{Dressel12}
\bibinfo{author}{\bibfnamefont{M.}~\bibnamefont{Dressel}},
  \bibinfo{author}{\bibfnamefont{M.}~\bibnamefont{Dumm}},
  \bibinfo{author}{\bibfnamefont{T.}~\bibnamefont{Knoblauch}},
  \bibinfo{author}{\bibfnamefont{B.}~\bibnamefont{K\"{o}hler}},
  \bibinfo{author}{\bibfnamefont{B.}~\bibnamefont{Salameh}}, \bibnamefont{and}
  \bibinfo{author}{\bibfnamefont{S.}~\bibnamefont{Yasin}},
  \bibinfo{journal}{Advances in Condensed Matter Physics}
  \textbf{\bibinfo{volume}{2012}}, \bibinfo{pages}{398721}
  (\bibinfo{year}{2012}),
  \urlprefix\url{http://dx.doi.org/10.1155/2012/398721}.

\bibitem[{\citenamefont{Torrance
  et~al.}(1981{\natexlab{a}})\citenamefont{Torrance, Girlando, Mayerle,
  Crowley, Lee, Batail, and LaPlaca}}]{Torrance81}
\bibinfo{author}{\bibfnamefont{J.~B.} \bibnamefont{Torrance}},
  \bibinfo{author}{\bibfnamefont{A.}~\bibnamefont{Girlando}},
  \bibinfo{author}{\bibfnamefont{J.~J.} \bibnamefont{Mayerle}},
  \bibinfo{author}{\bibfnamefont{J.~I.} \bibnamefont{Crowley}},
  \bibinfo{author}{\bibfnamefont{V.~Y.} \bibnamefont{Lee}},
  \bibinfo{author}{\bibfnamefont{P.}~\bibnamefont{Batail}}, \bibnamefont{and}
  \bibinfo{author}{\bibfnamefont{S.~J.} \bibnamefont{LaPlaca}},
  \bibinfo{journal}{Phys. Rev. Lett.} \textbf{\bibinfo{volume}{47}},
  \bibinfo{pages}{1747} (\bibinfo{year}{1981}{\natexlab{a}}),
  \urlprefix\url{http://link.aps.org/doi/10.1103/PhysRevLett.47.1747}.

\bibitem[{\citenamefont{Okamoto et~al.}(1991)\citenamefont{Okamoto, Mitani,
  Tokura, Koshihara, Komatsu, Iwasa, Koda, and Saito}}]{Okamoto91}
\bibinfo{author}{\bibfnamefont{H.}~\bibnamefont{Okamoto}},
  \bibinfo{author}{\bibfnamefont{T.}~\bibnamefont{Mitani}},
  \bibinfo{author}{\bibfnamefont{Y.}~\bibnamefont{Tokura}},
  \bibinfo{author}{\bibfnamefont{S.}~\bibnamefont{Koshihara}},
  \bibinfo{author}{\bibfnamefont{T.}~\bibnamefont{Komatsu}},
  \bibinfo{author}{\bibfnamefont{Y.}~\bibnamefont{Iwasa}},
  \bibinfo{author}{\bibfnamefont{T.}~\bibnamefont{Koda}}, \bibnamefont{and}
  \bibinfo{author}{\bibfnamefont{G.}~\bibnamefont{Saito}},
  \bibinfo{journal}{Phys. Rev. B} \textbf{\bibinfo{volume}{43}},
  \bibinfo{pages}{8224} (\bibinfo{year}{1991}),
  \urlprefix\url{http://link.aps.org/doi/10.1103/PhysRevB.43.8224}.

\bibitem[{\citenamefont{Le~Cointe et~al.}(1995)\citenamefont{Le~Cointe,
  Lem\'ee-Cailleau, Cailleau, Toudic, Toupet, Heger, Moussa, Schweiss, Kraft,
  and Karl}}]{LeCointe95}
\bibinfo{author}{\bibfnamefont{M.}~\bibnamefont{Le~Cointe}},
  \bibinfo{author}{\bibfnamefont{M.~H.} \bibnamefont{Lem\'ee-Cailleau}},
  \bibinfo{author}{\bibfnamefont{H.}~\bibnamefont{Cailleau}},
  \bibinfo{author}{\bibfnamefont{B.}~\bibnamefont{Toudic}},
  \bibinfo{author}{\bibfnamefont{L.}~\bibnamefont{Toupet}},
  \bibinfo{author}{\bibfnamefont{G.}~\bibnamefont{Heger}},
  \bibinfo{author}{\bibfnamefont{F.}~\bibnamefont{Moussa}},
  \bibinfo{author}{\bibfnamefont{P.}~\bibnamefont{Schweiss}},
  \bibinfo{author}{\bibfnamefont{K.~H.} \bibnamefont{Kraft}}, \bibnamefont{and}
  \bibinfo{author}{\bibfnamefont{N.}~\bibnamefont{Karl}},
  \bibinfo{journal}{Phys. Rev. B} \textbf{\bibinfo{volume}{51}},
  \bibinfo{pages}{3374} (\bibinfo{year}{1995}),
  \urlprefix\url{http://link.aps.org/doi/10.1103/PhysRevB.51.3374}.

\bibitem[{\citenamefont{Lem\'ee-Cailleau
  et~al.}(1997)\citenamefont{Lem\'ee-Cailleau, Le~Cointe, Cailleau, Luty,
  Moussa, Roos, Brinkmann, Toudic, Ayache, and Karl}}]{LemeeCailleau97}
\bibinfo{author}{\bibfnamefont{M.~H.} \bibnamefont{Lem\'ee-Cailleau}},
  \bibinfo{author}{\bibfnamefont{M.}~\bibnamefont{Le~Cointe}},
  \bibinfo{author}{\bibfnamefont{H.}~\bibnamefont{Cailleau}},
  \bibinfo{author}{\bibfnamefont{T.}~\bibnamefont{Luty}},
  \bibinfo{author}{\bibfnamefont{F.}~\bibnamefont{Moussa}},
  \bibinfo{author}{\bibfnamefont{J.}~\bibnamefont{Roos}},
  \bibinfo{author}{\bibfnamefont{D.}~\bibnamefont{Brinkmann}},
  \bibinfo{author}{\bibfnamefont{B.}~\bibnamefont{Toudic}},
  \bibinfo{author}{\bibfnamefont{C.}~\bibnamefont{Ayache}}, \bibnamefont{and}
  \bibinfo{author}{\bibfnamefont{N.}~\bibnamefont{Karl}},
  \bibinfo{journal}{Phys. Rev. Lett.} \textbf{\bibinfo{volume}{79}},
  \bibinfo{pages}{1690} (\bibinfo{year}{1997}),
  \urlprefix\url{http://link.aps.org/doi/10.1103/PhysRevLett.79.1690}.

\bibitem[{\citenamefont{Kishida et~al.}(2009)\citenamefont{Kishida, Takamatsu,
  Fujinuma, and Okamoto}}]{Kishida09}
\bibinfo{author}{\bibfnamefont{H.}~\bibnamefont{Kishida}},
  \bibinfo{author}{\bibfnamefont{H.}~\bibnamefont{Takamatsu}},
  \bibinfo{author}{\bibfnamefont{K.}~\bibnamefont{Fujinuma}}, \bibnamefont{and}
  \bibinfo{author}{\bibfnamefont{H.}~\bibnamefont{Okamoto}},
  \bibinfo{journal}{Phys. Rev. B} \textbf{\bibinfo{volume}{80}},
  \bibinfo{pages}{205201} (\bibinfo{year}{2009}),
  \urlprefix\url{http://link.aps.org/doi/10.1103/PhysRevB.80.205201}.

\bibitem[{\citenamefont{Masino et~al.}(2006)\citenamefont{Masino, Girlando,
  Brillante, Valle, Venuti, Drichko, and Dressel}}]{Masino200671}
\bibinfo{author}{\bibfnamefont{M.}~\bibnamefont{Masino}},
  \bibinfo{author}{\bibfnamefont{A.}~\bibnamefont{Girlando}},
  \bibinfo{author}{\bibfnamefont{A.}~\bibnamefont{Brillante}},
  \bibinfo{author}{\bibfnamefont{R.~D.} \bibnamefont{Valle}},
  \bibinfo{author}{\bibfnamefont{E.}~\bibnamefont{Venuti}},
  \bibinfo{author}{\bibfnamefont{N.}~\bibnamefont{Drichko}}, \bibnamefont{and}
  \bibinfo{author}{\bibfnamefont{M.}~\bibnamefont{Dressel}},
  \bibinfo{journal}{Chem. Phys.} \textbf{\bibinfo{volume}{325}},
  \bibinfo{pages}{71 } (\bibinfo{year}{2006}), ISSN \bibinfo{issn}{0301-0104},
  \urlprefix\url{http://www.sciencedirect.com/science/article/pii/S0301010405004428}.

\bibitem[{\citenamefont{Girlando et~al.}(2008)\citenamefont{Girlando, Masino,
  Painelli, Drichko, Dressel, Brillante, Della~Valle, and Venuti}}]{Girlando08}
\bibinfo{author}{\bibfnamefont{A.}~\bibnamefont{Girlando}},
  \bibinfo{author}{\bibfnamefont{M.}~\bibnamefont{Masino}},
  \bibinfo{author}{\bibfnamefont{A.}~\bibnamefont{Painelli}},
  \bibinfo{author}{\bibfnamefont{N.}~\bibnamefont{Drichko}},
  \bibinfo{author}{\bibfnamefont{M.}~\bibnamefont{Dressel}},
  \bibinfo{author}{\bibfnamefont{A.}~\bibnamefont{Brillante}},
  \bibinfo{author}{\bibfnamefont{R.~G.} \bibnamefont{Della~Valle}},
  \bibnamefont{and} \bibinfo{author}{\bibfnamefont{E.}~\bibnamefont{Venuti}},
  \bibinfo{journal}{Phys. Rev. B} \textbf{\bibinfo{volume}{78}},
  \bibinfo{pages}{045103} (\bibinfo{year}{2008}),
  \urlprefix\url{http://link.aps.org/doi/10.1103/PhysRevB.78.045103}.

\bibitem[{\citenamefont{Nagahori et~al.}(2013)\citenamefont{Nagahori, Kubota,
  and Itoh}}]{Nagahori13}
\bibinfo{author}{\bibfnamefont{A.}~\bibnamefont{Nagahori}},
  \bibinfo{author}{\bibfnamefont{N.}~\bibnamefont{Kubota}}, \bibnamefont{and}
  \bibinfo{author}{\bibfnamefont{C.}~\bibnamefont{Itoh}},
  \bibinfo{journal}{Eur. Phys. J. B} \textbf{\bibinfo{volume}{86}},
  \bibinfo{pages}{1} (\bibinfo{year}{2013}), ISSN \bibinfo{issn}{1434-6028},
  \urlprefix\url{http://dx.doi.org/10.1140/epjb/e2013-30616-4}.

\bibitem[{\citenamefont{Tokura et~al.}(1988)\citenamefont{Tokura, Okamoto,
  Koda, Mitani, and Saito}}]{Tokura88}
\bibinfo{author}{\bibfnamefont{Y.}~\bibnamefont{Tokura}},
  \bibinfo{author}{\bibfnamefont{H.}~\bibnamefont{Okamoto}},
  \bibinfo{author}{\bibfnamefont{T.}~\bibnamefont{Koda}},
  \bibinfo{author}{\bibfnamefont{T.}~\bibnamefont{Mitani}}, \bibnamefont{and}
  \bibinfo{author}{\bibfnamefont{G.}~\bibnamefont{Saito}},
  \bibinfo{journal}{Phys. Rev. B} \textbf{\bibinfo{volume}{38}},
  \bibinfo{pages}{2215} (\bibinfo{year}{1988}),
  \urlprefix\url{http://link.aps.org/doi/10.1103/PhysRevB.38.2215}.

\bibitem[{\citenamefont{Kobayashi et~al.}(2012)\citenamefont{Kobayashi, Horiuchi, Kumai, Kagawa, Murakami, and Tokura}}]{Kobayashi12}
\bibinfo{author}{\bibfnamefont{K.}~\bibnamefont{Kobayashi}},
  \bibinfo{author}{\bibfnamefont{S.}~\bibnamefont{Horiuchi}},
  \bibinfo{author}{\bibfnamefont{R.}~\bibnamefont{Kumai}},
  \bibinfo{author}{\bibfnamefont{F.}~\bibnamefont{Kagawa}},
  \bibinfo{author}{\bibfnamefont{Y.}~\bibnamefont{Murakami}}, \bibnamefont{and}
  \bibinfo{author}{\bibfnamefont{Y.}~\bibnamefont{Tokura}},
  \bibinfo{journal}{Phys. Rev. Lett.} \textbf{\bibinfo{volume}{108}},
  \bibinfo{pages}{237601} (\bibinfo{year}{2012}),
  \urlprefix\url{http://link.aps.org/doi/10.1103/PhysRevLett.108.237601}.

\bibitem[{\citenamefont{Giovannetti et~al.}(2009)\citenamefont{Giovannetti, Kumar, Stroppa, van den Brink, and Picozzi}}]{Giovannetti09}
\bibinfo{author}{\bibfnamefont{G.}~\bibnamefont{Giovannetti}},
  \bibinfo{author}{\bibfnamefont{S.}~\bibnamefont{Kumar}},
  \bibinfo{author}{\bibfnamefont{A.}~\bibnamefont{Stroppa}},
  \bibinfo{author}{\bibfnamefont{J.}~\bibnamefont{van den Brink}}, \bibnamefont{and}
  \bibinfo{author}{\bibfnamefont{S.}~\bibnamefont{Picozzi}},
  \bibinfo{journal}{Phys. Rev. Lett.} \textbf{\bibinfo{volume}{103}},
  \bibinfo{pages}{266401} (\bibinfo{year}{2009}),
  \urlprefix\url{http://link.aps.org/doi/10.1103/PhysRevLett.103.266401}.

\bibitem[{\citenamefont{Masino et~al.}(2007)\citenamefont{Masino, Girlando, and
  Brillante}}]{Masino07}
\bibinfo{author}{\bibfnamefont{M.}~\bibnamefont{Masino}},
  \bibinfo{author}{\bibfnamefont{A.}~\bibnamefont{Girlando}}, \bibnamefont{and}
  \bibinfo{author}{\bibfnamefont{A.}~\bibnamefont{Brillante}},
  \bibinfo{journal}{Phys. Rev. B} \textbf{\bibinfo{volume}{76}},
  \bibinfo{pages}{064114} (\bibinfo{year}{2007}),
  \urlprefix\url{http://link.aps.org/doi/10.1103/PhysRevB.76.064114}.

\bibitem[{\citenamefont{T. et~al.}(1998)\citenamefont{T., K., and
  G.}}]{Ishiguro98}
\bibinfo{author}{\bibfnamefont{T.}~\bibnamefont{Ishiguro}},
  \bibinfo{author}{\bibfnamefont{K.}~\bibnamefont{Yamaji}}, \bibnamefont{and}
  \bibinfo{author}{\bibfnamefont{G.}~\bibnamefont{Saito}},
  \emph{\bibinfo{title}{Organic Superconductors}} (\bibinfo{publisher}{Springer
  Verlag}, \bibinfo{address}{Berlin, Heidelberg, New York},
  \bibinfo{year}{1998}).

\bibitem[{\citenamefont{Torrance
  et~al.}(1981{\natexlab{b}})\citenamefont{Torrance, Vazquez, Mayerle, and
  Lee}}]{Torrance81PRL}
\bibinfo{author}{\bibfnamefont{J.~B.} \bibnamefont{Torrance}},
  \bibinfo{author}{\bibfnamefont{J.~E.} \bibnamefont{Vazquez}},
  \bibinfo{author}{\bibfnamefont{J.~J.} \bibnamefont{Mayerle}},
  \bibnamefont{and} \bibinfo{author}{\bibfnamefont{V.~Y.} \bibnamefont{Lee}},
  \bibinfo{journal}{Phys. Rev. Lett.} \textbf{\bibinfo{volume}{46}},
  \bibinfo{pages}{253} (\bibinfo{year}{1981}{\natexlab{b}}),
  \urlprefix\url{http://link.aps.org/doi/10.1103/PhysRevLett.46.253}.

\bibitem[{\citenamefont{Takaoka et~al.}(1987)\citenamefont{Takaoka, Kaneko,
  Okamoto, Tokura, Koda, Mitani, and Saito}}]{Takaoka87}
\bibinfo{author}{\bibfnamefont{K.}~\bibnamefont{Takaoka}},
  \bibinfo{author}{\bibfnamefont{Y.}~\bibnamefont{Kaneko}},
  \bibinfo{author}{\bibfnamefont{H.}~\bibnamefont{Okamoto}},
  \bibinfo{author}{\bibfnamefont{Y.}~\bibnamefont{Tokura}},
  \bibinfo{author}{\bibfnamefont{T.}~\bibnamefont{Koda}},
  \bibinfo{author}{\bibfnamefont{T.}~\bibnamefont{Mitani}}, \bibnamefont{and}
  \bibinfo{author}{\bibfnamefont{G.}~\bibnamefont{Saito}},
  \bibinfo{journal}{Phys. Rev. B} \textbf{\bibinfo{volume}{36}},
  \bibinfo{pages}{3884} (\bibinfo{year}{1987}),
  \urlprefix\url{http://link.aps.org/doi/10.1103/PhysRevB.36.3884}.

\bibitem[{\citenamefont{Moreac et~al.}(1996)\citenamefont{Moreac, Girard, and
  Delugeard}}]{Moreac96}
\bibinfo{author}{\bibfnamefont{A.}~\bibnamefont{Moreac}},
  \bibinfo{author}{\bibfnamefont{A.}~\bibnamefont{Girard}}, \bibnamefont{and}
  \bibinfo{author}{\bibfnamefont{Y.}~\bibnamefont{Delugeard}},
  \bibinfo{journal}{J. Phys.: Condens. Matter}
  \textbf{\bibinfo{volume}{8}}, \bibinfo{pages}{3569} (\bibinfo{year}{1996}),
  \urlprefix\url{http://stacks.iop.org/0953-8984/8/i=20/a=005}.

\bibitem[{\citenamefont{Tokura et~al.}(1986)\citenamefont{Tokura, Okamoto,
  Koda, and Mitani}}]{Tokura86}
\bibinfo{author}{\bibfnamefont{Y.}~\bibnamefont{Tokura}},
  \bibinfo{author}{\bibfnamefont{H.}~\bibnamefont{Okamoto}},
  \bibinfo{author}{\bibfnamefont{T.}~\bibnamefont{Koda}}, \bibnamefont{and}
  \bibinfo{author}{\bibfnamefont{T.}~\bibnamefont{Mitani}},
  \bibinfo{journal}{Solid State Commun.} \textbf{\bibinfo{volume}{57}},
  \bibinfo{pages}{607 } (\bibinfo{year}{1986}), ISSN \bibinfo{issn}{0038-1098},
  \urlprefix\url{http://www.sciencedirect.com/science/article/pii/0038109886903327}.

\bibitem[{\citenamefont{Mitani et~al.}(1987)\citenamefont{Mitani, Kaneko,
  Tanuma, Tokura, Koda, and Saito}}]{Mitani87}
\bibinfo{author}{\bibfnamefont{T.}~\bibnamefont{Mitani}},
  \bibinfo{author}{\bibfnamefont{Y.}~\bibnamefont{Kaneko}},
  \bibinfo{author}{\bibfnamefont{S.}~\bibnamefont{Tanuma}},
  \bibinfo{author}{\bibfnamefont{Y.}~\bibnamefont{Tokura}},
  \bibinfo{author}{\bibfnamefont{T.}~\bibnamefont{Koda}}, \bibnamefont{and}
  \bibinfo{author}{\bibfnamefont{G.}~\bibnamefont{Saito}},
  \bibinfo{journal}{Phys. Rev. B} \textbf{\bibinfo{volume}{35}},
  \bibinfo{pages}{427} (\bibinfo{year}{1987}),
  \urlprefix\url{http://link.aps.org/doi/10.1103/PhysRevB.35.427}.

\bibitem[{\citenamefont{Matsuzaki et~al.}(2005)\citenamefont{Matsuzaki,
  Takamatsu, Kishida, and Okamoto}}]{Matsuzaki05}
\bibinfo{author}{\bibfnamefont{H.}~\bibnamefont{Matsuzaki}},
  \bibinfo{author}{\bibfnamefont{H.}~\bibnamefont{Takamatsu}},
  \bibinfo{author}{\bibfnamefont{H.}~\bibnamefont{Kishida}}, \bibnamefont{and}
  \bibinfo{author}{\bibfnamefont{H.}~\bibnamefont{Okamoto}},
  \bibinfo{journal}{J. Phys. Soc. Jpn.}
  \textbf{\bibinfo{volume}{74}}, \bibinfo{pages}{2925} (\bibinfo{year}{2005}),
  \urlprefix\url{http://jpsj.ipap.jp/link?JPSJ/74/2925/}.

\bibitem[{\citenamefont{Dressel and Drichko}(2004)}]{Dressel04}
\bibinfo{author}{\bibfnamefont{M.}~\bibnamefont{Dressel}} \bibnamefont{and}
  \bibinfo{author}{\bibfnamefont{N.}~\bibnamefont{Drichko}},
  \bibinfo{journal}{Chem. Rev.} \textbf{\bibinfo{volume}{104}},
  \bibinfo{pages}{5689} (\bibinfo{year}{2004}),
  \urlprefix\url{http://pubs.acs.org/doi/abs/10.1021/cr030642f}.

\bibitem[{\citenamefont{Drichko et~al.}(2009)\citenamefont{Drichko, Kaiser,
  Sun, Clauss, Dressel, Mori, Schlueter, Zhyliaeva, Torunova, and
  Lyubovskaya}}]{Drichko09}
\bibinfo{author}{\bibfnamefont{N.}~\bibnamefont{Drichko}},
  \bibinfo{author}{\bibfnamefont{S.}~\bibnamefont{Kaiser}},
  \bibinfo{author}{\bibfnamefont{Y.}~\bibnamefont{Sun}},
  \bibinfo{author}{\bibfnamefont{C.}~\bibnamefont{Clauss}},
  \bibinfo{author}{\bibfnamefont{M.}~\bibnamefont{Dressel}},
  \bibinfo{author}{\bibfnamefont{H.}~\bibnamefont{Mori}},
  \bibinfo{author}{\bibfnamefont{J.}~\bibnamefont{Schlueter}},
  \bibinfo{author}{\bibfnamefont{E.~I.} \bibnamefont{Zhyliaeva}},
  \bibinfo{author}{\bibfnamefont{S.~A.} \bibnamefont{Torunova}},
  \bibnamefont{and} \bibinfo{author}{\bibfnamefont{R.~N.}
  \bibnamefont{Lyubovskaya}}, \bibinfo{journal}{Physica B}
  \textbf{\bibinfo{volume}{404}}, \bibinfo{pages}{490 } (\bibinfo{year}{2009}),
  ISSN \bibinfo{issn}{0921-4526},
  \urlprefix\url{http://www.sciencedirect.com/science/article/pii/S0921452608005681}.

\bibitem[{\citenamefont{Girlando}(2011)}]{Girlando11}
\bibinfo{author}{\bibfnamefont{A.}~\bibnamefont{Girlando}},
  \bibinfo{journal}{J. Phys. Chem. C}
  \textbf{\bibinfo{volume}{115}}, \bibinfo{pages}{19371}
  (\bibinfo{year}{2011}),
  \eprint{http://pubs.acs.org/doi/pdf/10.1021/jp206171r},
  \urlprefix\url{http://pubs.acs.org/doi/abs/10.1021/jp206171r}.

\bibitem[{\citenamefont{Sedlmeier et~al.}(2012)\citenamefont{Sedlmeier,
  Els\"{a}sser, Neubauer, Beyer, Wu, Ivek, Tomi\'{c}, Schlueter, and
  Dressel}}]{Sedlmeier12}
\bibinfo{author}{\bibfnamefont{K.}~\bibnamefont{Sedlmeier}},
  \bibinfo{author}{\bibfnamefont{S.}~\bibnamefont{Els\"{a}sser}},
  \bibinfo{author}{\bibfnamefont{D.}~\bibnamefont{Neubauer}},
  \bibinfo{author}{\bibfnamefont{R.}~\bibnamefont{Beyer}},
  \bibinfo{author}{\bibfnamefont{D.}~\bibnamefont{Wu}},
  \bibinfo{author}{\bibfnamefont{T.}~\bibnamefont{Ivek}},
  \bibinfo{author}{\bibfnamefont{S.}~\bibnamefont{Tomi\'{c}}},
  \bibinfo{author}{\bibfnamefont{J.~A.} \bibnamefont{Schlueter}},
  \bibnamefont{and} \bibinfo{author}{\bibfnamefont{M.}~\bibnamefont{Dressel}},
  \bibinfo{journal}{Phys. Rev. B} \textbf{\bibinfo{volume}{86}},
  \bibinfo{pages}{245103} (\bibinfo{year}{2012}),
  \urlprefix\url{http://link.aps.org/doi/10.1103/PhysRevB.86.245103}.

\bibitem[{\citenamefont{Girlando et~al.}(1983)\citenamefont{Girlando, Marzola,
  Pecile, and Torrance}}]{Girlando83}
\bibinfo{author}{\bibfnamefont{A.}~\bibnamefont{Girlando}},
  \bibinfo{author}{\bibfnamefont{F.}~\bibnamefont{Marzola}},
  \bibinfo{author}{\bibfnamefont{C.}~\bibnamefont{Pecile}}, \bibnamefont{and}
  \bibinfo{author}{\bibfnamefont{J.~B.} \bibnamefont{Torrance}},
  \bibinfo{journal}{J. Chem. Phys.}
  \textbf{\bibinfo{volume}{79}}, \bibinfo{pages}{1075} (\bibinfo{year}{1983}),
  \urlprefix\url{http://link.aip.org/link/?JCP/79/1075/1}.

\bibitem[{\citenamefont{Masino et~al.}(2004)\citenamefont{Masino, Girlando,
  Brillante, Valle, and Venuti}}]{Masino04}
\bibinfo{author}{\bibfnamefont{M.}~\bibnamefont{Masino}},
  \bibinfo{author}{\bibfnamefont{A.}~\bibnamefont{Girlando}},
  \bibinfo{author}{\bibfnamefont{A.}~\bibnamefont{Brillante}},
  \bibinfo{author}{\bibfnamefont{R.~D.} \bibnamefont{Valle}}, \bibnamefont{and}
  \bibinfo{author}{\bibfnamefont{E.}~\bibnamefont{Venuti}},
  \bibinfo{journal}{Mater. Sci.-Poland} \textbf{\bibinfo{volume}{22}}
  (\bibinfo{year}{2004}),
  \urlprefix\url{http://materialsscience.pwr.wroc.pl/index.php?id=5&vol=vol22no4}.

\bibitem[{\citenamefont{Beyer et~al.}(2014)\citenamefont{Beyer, Bari\v{s}i\'c,
  and Dressel}}]{Beyer}
\bibinfo{author}{\bibfnamefont{R.}~\bibnamefont{Beyer}},
  \bibinfo{author}{\bibfnamefont{N.}~\bibnamefont{Bari\v{s}i\'c}},
  \bibnamefont{and} \bibinfo{author}{\bibfnamefont{M.}~\bibnamefont{Dressel}},
  \bibinfo{journal}{to be published}  (\bibinfo{year}{2014}).

\bibitem[{\citenamefont{Demianets et~al.}(1980)\citenamefont{Demianets,
  Emelchenko, Hesse, Karl, Lobachev, and Maier}}]{Demianets80}
\bibinfo{author}{\bibfnamefont{L.~N.} \bibnamefont{Demianets}},
  \bibinfo{author}{\bibfnamefont{G.~A.} \bibnamefont{Emelchenko}},
  \bibinfo{author}{\bibfnamefont{J.}~\bibnamefont{Hesse}},
  \bibinfo{author}{\bibfnamefont{N.}~\bibnamefont{Karl}},
  \bibinfo{author}{\bibfnamefont{A.~N.} \bibnamefont{Lobachev}},
  \bibnamefont{and} \bibinfo{author}{\bibfnamefont{H.}~\bibnamefont{Maier}},
  \emph{\bibinfo{title}{Organic Crystals, Germanates, Semiconductors}}
  (\bibinfo{publisher}{Springer}, \bibinfo{address}{Berlin Heidelberg},
  \bibinfo{year}{1980}),
  \urlprefix\url{http://link.springer.com/book/10.1007%2F978-3-642-67764-9}.

\bibitem[{\citenamefont{Dressel and Gr{\"u}ner}(2002)}]{DresselGruner02}
\bibinfo{author}{\bibfnamefont{M.}~\bibnamefont{Dressel}} \bibnamefont{and}
  \bibinfo{author}{\bibfnamefont{G.}~\bibnamefont{Gr{\"u}ner}},
  \emph{\bibinfo{title}{Electrodynamics of Solids: Optical Properties of
  Electrons in Matter}} (\bibinfo{publisher}{Cambridge University Press},
  \bibinfo{address}{Cambridge}, \bibinfo{year}{2002}).

\bibitem[{\citenamefont{Fano}(1961)}]{Fano61}
\bibinfo{author}{\bibfnamefont{U.}~\bibnamefont{Fano}}, \bibinfo{journal}{Phys.
  Rev.} \textbf{\bibinfo{volume}{124}}, \bibinfo{pages}{1866}
  (\bibinfo{year}{1961}),
  \urlprefix\url{http://link.aps.org/doi/10.1103/PhysRev.124.1866}.

\bibitem[{\citenamefont{Damascelli et~al.}(1997)\citenamefont{Damascelli,
  Schulte, van~der Marel, and Menovsky}}]{Damascelli97}
\bibinfo{author}{\bibfnamefont{A.}~\bibnamefont{Damascelli}},
  \bibinfo{author}{\bibfnamefont{K.}~\bibnamefont{Schulte}},
  \bibinfo{author}{\bibfnamefont{D.}~\bibnamefont{van~der Marel}},
  \bibnamefont{and} \bibinfo{author}{\bibfnamefont{A.~A.}
  \bibnamefont{Menovsky}}, \bibinfo{journal}{Phys. Rev. B}
  \textbf{\bibinfo{volume}{55}}, \bibinfo{pages}{R4863} (\bibinfo{year}{1997}),
  \urlprefix\url{http://link.aps.org/doi/10.1103/PhysRevB.55.R4863}.

\bibitem[{\citenamefont{Lemee-Cailleau
  et~al.}(1992)\citenamefont{Lemee-Cailleau, Toudic, Cailleau, Moussa, Cointe,
  Silly, and Karl}}]{LemeeCailleau92}
\bibinfo{author}{\bibfnamefont{M.~H.} \bibnamefont{Lemee-Cailleau}},
  \bibinfo{author}{\bibfnamefont{B.}~\bibnamefont{Toudic}},
  \bibinfo{author}{\bibfnamefont{H.}~\bibnamefont{Cailleau}},
  \bibinfo{author}{\bibfnamefont{F.}~\bibnamefont{Moussa}},
  \bibinfo{author}{\bibfnamefont{M.~L.} \bibnamefont{Cointe}},
  \bibinfo{author}{\bibfnamefont{G.}~\bibnamefont{Silly}}, \bibnamefont{and}
  \bibinfo{author}{\bibfnamefont{N.}~\bibnamefont{Karl}},
  \bibinfo{journal}{Ferroelectrics} \textbf{\bibinfo{volume}{127}},
  \bibinfo{pages}{19} (\bibinfo{year}{1992}),
  \urlprefix\url{http://dx.doi.org/10.1080/00150199208223340}.

\end{thebibliography}

\end{document}